\documentclass[twocolumn]{aastex63}
\usepackage{natbib,amssymb,amsmath,graphicx,here}

\shorttitle{$\kappa$ And b with ALES}
\shortauthors{Stone et al.}

\begin{document}

\title{High Contrast Thermal Infrared Spectroscopy with ALES: The 3-4$\mu$m Spectrum of $\kappa$ Andromedae b}

\correspondingauthor{Jordan M Stone}
\email{jordan.stone@nrl.navy.mil}

\author[0000-0003-0454-3718]{Jordan M. Stone}
\altaffiliation{Hubble Fellow} 
\affiliation{Naval Research Laboratory, 
Remote Sensing Division, 
4555 Overlook Ave SW, 
Washington, DC 20375 USA} 
\affiliation{Steward Observatory,
University of Arizona,
933 N. Cherry Ave,
Tucson, AZ 85721-0065 USA} 

\author{Travis Barman}
\affiliation{Lunar and Planetary Laboratory, 
The University of Arizona, 
1629 E. Univ. Blvd., 
Tucson, AZ 85721 USA}

\author{Andrew J. Skemer}
\affiliation{Department of Astronomy and Astrophysics, 
University of California, Santa Cruz, 
1156 High St, 
Santa Cruz, CA 95064, USA}

\author[0000-0002-1764-2494]{Zackery W. Briesemeister}
\affiliation{Department of Astronomy and Astrophysics, 
University of California, Santa Cruz, 
1156 High St, 
Santa Cruz, CA 95064, USA}

\author{Laci S. Brock}
\affiliation{Lunar and Planetary Laboratory, 
The University of Arizona, 
1629 E. Univ. Blvd., 
Tucson, AZ 85721 USA}

\author{Philip M. Hinz}
\affiliation{Department of Astronomy and Astrophysics, 
University of California, Santa Cruz, 
1156 High St, 
Santa Cruz, CA 95064, USA}

\author{Jarron M. Leisenring}
\affiliation{Steward Observatory,
University of Arizona,
933 N. Cherry Ave,
Tucson, AZ 85721-0065 USA} 

\author[0000-0001-6567-627X]{Charles E. Woodward}
\affiliation{Minnesota Institute of Astrophysics, 
University of Minnesota,
116 Church Street, SE,
Minneapolis, MN 55455, USA}

\author{Michael F. Skrutskie}
\affiliation{Department of Astronomy, University of Virginia, Charlottesville,
VA 22904, USA}

\author[0000-0003-3819-0076]{Eckhart Spalding}
\affiliation{Steward Observatory,
University of Arizona,
933 N. Cherry Ave,
Tucson, AZ 85721-0065 USA} 

\begin{abstract} 

We present the first $L-$band (2.8 to 4.1~$\mu$m) spectroscopy of
$\kappa$~Andromedae~b, a $\sim20~M_{\mathrm{Jup}}$ companion orbiting at
$1\arcsec$ projected separation from its B9-type stellar host. We combine our
Large Binocular Telescope ALES integral field spectrograph data with
measurements from other instruments to analyze the atmosphere and physical
characteristics of $\kappa$~And~b. We report a discrepancy of $\sim20\%$
($2\sigma$) in the $L^{\prime}$ flux of $\kappa$~And~b when comparing to
previously published values. We add an additional $L^{\prime}$ constraint using
an unpublished imaging dataset collected in 2013 using LBTI/LMIRCam, the
instrument in which the ALES module has been built.  The LMIRCam measurement is
consistent with the ALES measurement, both suggesting a fainter $L$-band
scaling than previous studies. The data, assuming the flux scaling measured by
ALES and LMIRCam imaging, are well fit by an L3-type brown dwarf.  Atmospheric
model fits to measurements spanning 0.9-4.8~$\mu$m reveal some tension with the
predictions of evolutionary models, but the proper choice of cloud parameters
can provide some relief. In particular, models with clouds extending to
very-low pressures composed of grains $\leq1~\mu$m appear to be necessary. If
the brighter $L^{\prime}$ photometry is accurate, there is a hint that
sub-solar metallicity may be required.  \end{abstract}


\section{Introduction} 

The $\kappa$~And system consists of a late B-type star orbited by a substellar
companion at $\sim1\arcsec$ projected separation
\citep[$\sim50$~au,][]{Carson2013}. The mass of the companion,
$\sim20~M_{\mathrm{Jup}}$ \citep[e.g.,][]{Uyama2020} is estimated by combining
evolutionary models \citep[e.g.,][]{Baraffe2015}, constraints on the bolometric
luminosity \citep[this work,][]{Bonnefoy2014}, and an age estimate for the
system \citep{Bonnefoy2014, Jones2016}.  

Age dating early-type stars like $\kappa$~And is challenging. However, if the
early-type star is a member of a kinematic association that has later type
stars whose ages can be more definitively determined, more constrained age
estimates are possible.  The kinematics of the $\kappa$~And system are
suggestive of membership in the Columba young association \citep{Zuckerman2011,
Carson2013, Bonnefoy2014}, which has an age of $42^{+6}_{-4}$~Myr
\citep{Bell2015}.  \citet{Bonnefoy2014} report a 95\% chance that $\kappa$~And
is part of the Columba young association \citep[using the online tool reported
in][]{Malo2013}. Using the same analysis, \citet{Bonnefoy2014} report a 98\%
chance that the four-planet HR~8799 system is a member of Columba, implying the
two systems are siblings. However, using astrometric constraints provided by
$Gaia$ DR2 \citep{GaiaCollaboration2018} and the online BANYAN
$\Sigma$\footnote{http://www.exoplanetes.umontreal.ca/banyan/banyansigma.php}
tool \citep[which supercedes the Malo et al. (2013) version,][]{Gagne2018},
there is a 31\% chance that $\kappa$~And is a member of Columba and a 42\%
chance for HR~8799.  Yet, youth, and membership in the young association, is
supported for both objects by comparison to evolutionary models. This is
typically an imprecise exercise for early type stars, but both stars have had
their photospheres resolved with long-baseline optical interferometry, meaning
their fundamental parameters are particularly well constrained and model
comparisons more precise \citep{Baines2012, Jones2016}. For $\kappa$~And the
age constraint from evolutionary models is $47^{+27}_{-40}$~Myr, consistent
with membership in Columba.

While the planetary nature of the HR~8799 companions is secure ---masses
estimated to be below the deuterium burning limit, system architecture
comprising co-planar non-hierarchical orbits--- the nature of $\kappa$~And~b is
more uncertain. On the one hand, the estimated mass of $\kappa$~And~b is larger
than the deuterium burning limit.  On the other hand, the mass ratio with
$\kappa$~And~A \citep[$M_{\mathrm{A}}=2.8~M_{\odot}$,][]{Jones2016} is almost
identical to the star/planet mass ratios ($\sim0.7\%$) within the HR~8799
system. Consequently, the $\kappa$~And system provides an excellent target to
study the formation of objects in the low-mass brown dwarf/high-mass planet
regime where objects of similar mass can have distinct formation histories
\citep[e.g.,][]{Brandt2014, Reggiani2016, Wagner2019}.

The eccentricity of the $\kappa$~And~b orbit, recently refined by
\citet{Uyama2020}, is more likely to be drawn from the brown dwarf population
than the planetary population according to recent work by \citet{Bowler2020}.
However, the system is part of the sample that \citet{Bowler2020} used to
constrain the eccentricity distributions. When constructing eccentricity
distributions using mass ratio bins rather than mass bins to define the
populations, \citet{Bowler2020} include $\kappa$~And~b in the planetary group
and the distinction between the eccentricity distributions becomes more
ambiguous.

\begin{figure}
\includegraphics[width=\linewidth]{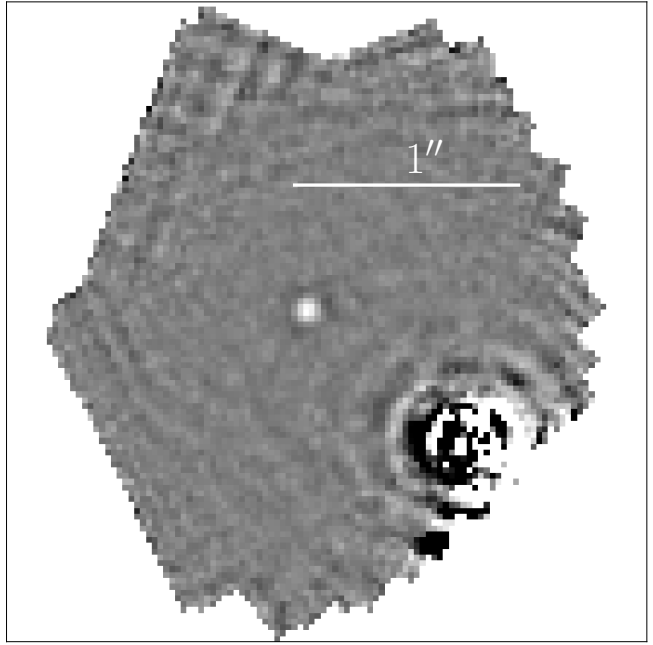}
\caption{ALES image of $\kappa$~And~b made by summing the (x,y,$\lambda$) data
cube over the wavelength dimension. \label{cubestack}}
\end{figure}

Atmospheres provide an avenue for constraining companion composition, which can
help distinguish formation processes. For example, the binary star formation
mechanism results in similar composition for both components
\citep{Desidera2006}, but the planet formation process can naturally enhance
the metallicity of a planet over that of the host star
\citep[e.g.,][]{Pollack1986, Wong2004, Boley2011}. As a result, differential
metallicity is one potential route to distinguish extreme mass ratio binaries
from planetary systems. However, the number of important atmospheric parameters
to constrain is large for young low-mass objects \citep[e.g.,][]{Barman2011a,
Skemer2011, Marley2012} making metallicity constraints challenging due to model
degeneracies connecting temperature, gravity, chemistry, and cloud properties
\citep[e.g.,][]{Rajan2017}. Overcoming these degeneracies is particularly
challenging due to the complex nature of clouds and atmospheric condensates.

Breaking these degeneracies benefits from measurements covering a wide range of
wavelengths \citep[e.g.,][]{Stephens2009, Morzinski2015, Skemer2016}.  In this paper we
report the first $L$ band (2.8-4.1~$\mu$m) spectroscopy of $\kappa$~And~b,
harnessing the new thermal-infrared integral field spectroscopy capabilities
delivered by the Arizona Lenslets for Exoplanet Spectroscopy (ALES) instrument
embedded within LMIRCam \citep{Skrutskie2010, Leisenring2012} and operating
within the Large Binocular Telescope Interferometer architecture
\citep{Skemer2015, Skemer2018}.  We targeted $\kappa$~And~b with ALES to
increase the wavelength range over which spectroscopic measurements probe the
atmosphere of the object.  In Section \ref{ObsSec} we describe our
observations.  We discuss our data reduction approach in Section \ref{dataSec}.
In Section \ref{analysisSec} we present our results and describe our approach
to fitting model atmosphere spectra to ALES data combined with measurements
from other instruments spanning from 0.9 to 5~$\mu$m. In Section \ref{discSec}
we provide a discussion of our results focusing on the contribution to the fit
provided by ALES.

\begin{figure}
\includegraphics[width=\linewidth]{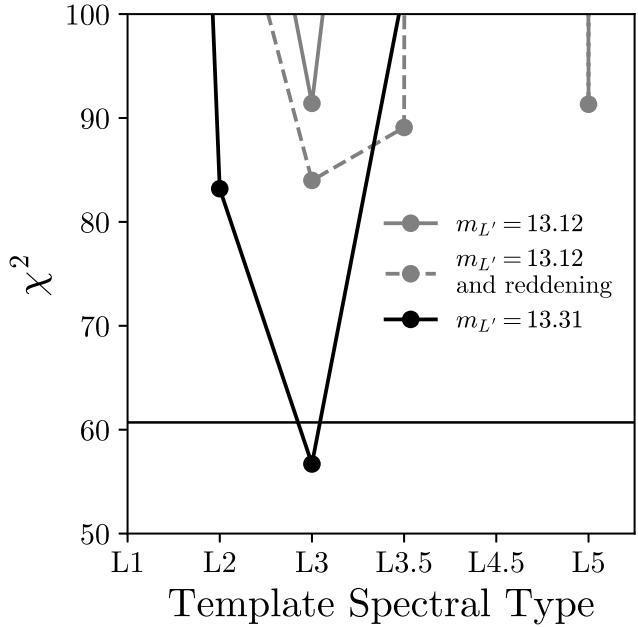}
\caption{$\chi^{2}$ as a function of the Spex Prism Library template spectrum
spectral type. The solid black curve applies to the case where the $L^{\prime}$
photometry of $\kappa$~And~b is consistent with ALES and LMIRCam imaging measurements.
The solid gray curve applies to the case where the $L$-band spectrum is scaled
to match the $L^{\prime}$ flux measured by \citet{Carson2013} and
\citet{Bonnefoy2014}. The dashed curve is like the gray, but allowing the
template spectra to be reddened according to the prescription of
\citet{Cardelli1989}.\label{SpexChi}} 
\end{figure}
\begin{figure*}
\includegraphics[width=\linewidth]{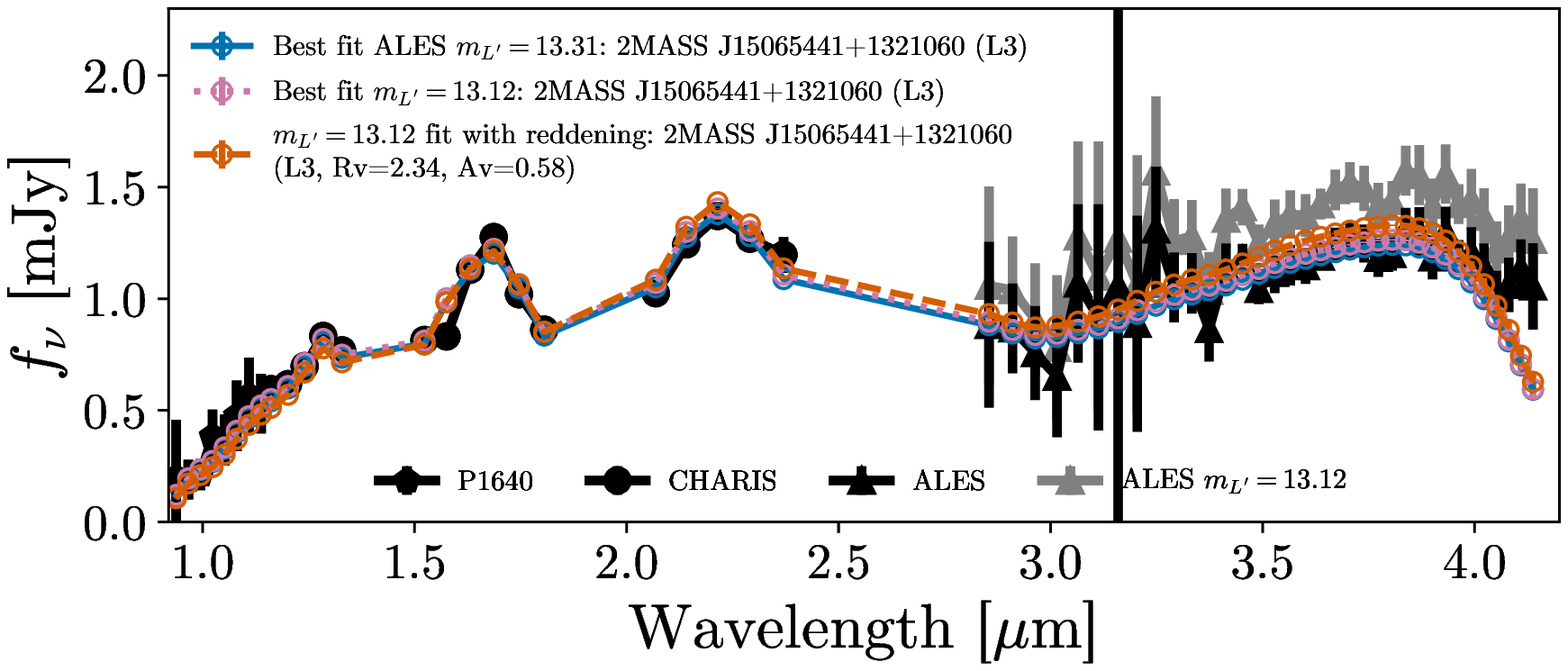}
\caption{Integral field spectroscopy of $\kappa$~And~b covering 0.9 to
4.1~$\mu$m compared to the L3-type field brown dwarf 2MASS~J15065441+1321060
  \citep{Cushing2008}.  The P1640 Y-band spectrum is from \citet{Hinkley2013},
and the CHARIS J-, H-, and K-band spectrum is from \citet{Currie2018}. Two
versions of the ALES spectrum are shown. In black, the spectrum with scaling
based on the ALES and LMIRCam measured photometry. In gray, the ALES spectrum
is shown scaled to be consistent with Subaru and Keck photometry
\citep{Carson2013, Bonnefoy2014}.  The L3 template is a good fit to the black
data points. While L3 is the best fit to the SED using the gray points, it is
not as good. Adding reddening, following \citet{Hiranaka2016}, cannot
significantly improve the fit to the data using the gray points because
reddening affects the well constrained J, H, K data more than the
L-band.\label{SpexSpec}} 
\end{figure*}

\section{Observations} \label{ObsSec} 
We observed the $\kappa$~Andromedae system on UT 2016 November 13, using
LBTI/ALES in its $2.8~\mu\mathrm{m} - 4.1~\mu\mathrm{m}$ mode with spectral
resolution $R\sim20$.  The median seeing was $1\arcsec$, varying between
$0\farcs8$ and $1\farcs2$. We used only the left side of the LBT aperture and
adaptive optics system, correcting 400 modes with the deformable secondary at
1kHz loop speed.  Thin cirrus were present and atmospheric transmission was
poor short of 3.4 $\mu\mathrm{m}$ due to telluric water-ice absorption.

During our observations, LBTI/ALES was in the process of a multi-step upgrade
and in an intermediate state, delivering a grid of $74\times86$ spaxels
covering a $1\farcs93\times2\farcs24$ field of view ($0\farcs026$
spaxel$^{-1}$). In this early implementation, ALES spaxels were affected by
strong off-axis astigmatism, delivering significantly worse data near the edges
of the field. The lenselet array has since been upgraded \citep{Skemer2018,
Hinz2018}.

We observed using a three-point nod pattern alternating from a position with
the primary star centered on the lenslet array (10 frames), to a position with
the companion centered on the array(60 frames), then to a nearby sky position
(60 frames). This approach ensured that $\kappa$ And b was always observed
through a region of high optical quality even as its position rotated in the
frame with the parallactic angle \citep{Stone2018}. We chose a 1.16~s exposure
time to keep the sky emission in the linear range of the HAWAII-2RG detector.
In all, we executed 16 three-point nod cycles, collecting 18.5 minutes of
exposure time with the companion positioned in the region of the ALES field of
view with low astigmatism.  These data include $138^{\degr}$ of parallactic
angle change. 

Immediately following our companion observations, we collected unsaturated
frames of the primary star using 0.58~s exposures. These data were used for
simultaneous telluric and photometric calibration, described below.  For
wavelength calibration, we observed a nearby sky position through four
narrowband ($R\sim100$) filters positioned upstream of the ALES optics within
LMIRCam \citep{Stone2018}. At the thermal-IR wavelengths where ALES operates the
blank sky provides plenty of flux through the narrowband filters so that
wavelength calibration can be carried out efficiently.

\section{Data Reduction} \label{dataSec}
\subsection{Making Datacubes} 
The detector used by ALES exhibits time-variable offsets within each 64-column
readout channel due to drifting biases in the readout amplifiers. We correct
for this by measuring the offset in each channel using a median of the pixels
in the first 20 rows ---the extent of the detector not covered by the ALES
lenslet array. We found that this approach provided a better correction
compared to using only the four overscan rows of the detector. After channel
offset correction, we corrected bad pixels by replacing them with the median of
their nearest four good neighbors. 

For each three-point nod cycle, we median combined the (10) primary, (60)
companion, and (60) sky frames. We then subtracted the median sky frame from
the median on-source images.  We extracted $(x, y, \lambda)$ data cubes using
an inverse variance and spatial profile weighted extraction approach on each of
the $74\times86$ micro-spectra across the ALES field \citep{Horne1986,
Briesemeister2018}. Since our lenslet array was paired with a pinhole grid to
alleviate crosstalk due to diffraction \citep{Skemer2015}, we noticed that the
spatial profile of each micro-spectrum was not a strong function of wavelength.
Consequently, to measure the spatial profile of each micro-spectrum, we first
collapsed each micro-spectrum along the wavelength axis to create a single high
signal-to-noise spatial profile which we then applied at all wavelengths.
A typical spatial profile had $\sim5$ HAWAII-2RG pixels full width at half
maximum. Our cubes included 38 wavelength slices spanning
$2.85~\mu\mathrm{m}$ to $4.19~\mu\mathrm{m}$. Resulting spectral image cubes
revealed a point-spread function (PSF) having 4.1 spatial pixels full width at
half maximum at 3.77~$\mu$m.

\begin{figure*}
\includegraphics[width=\linewidth]{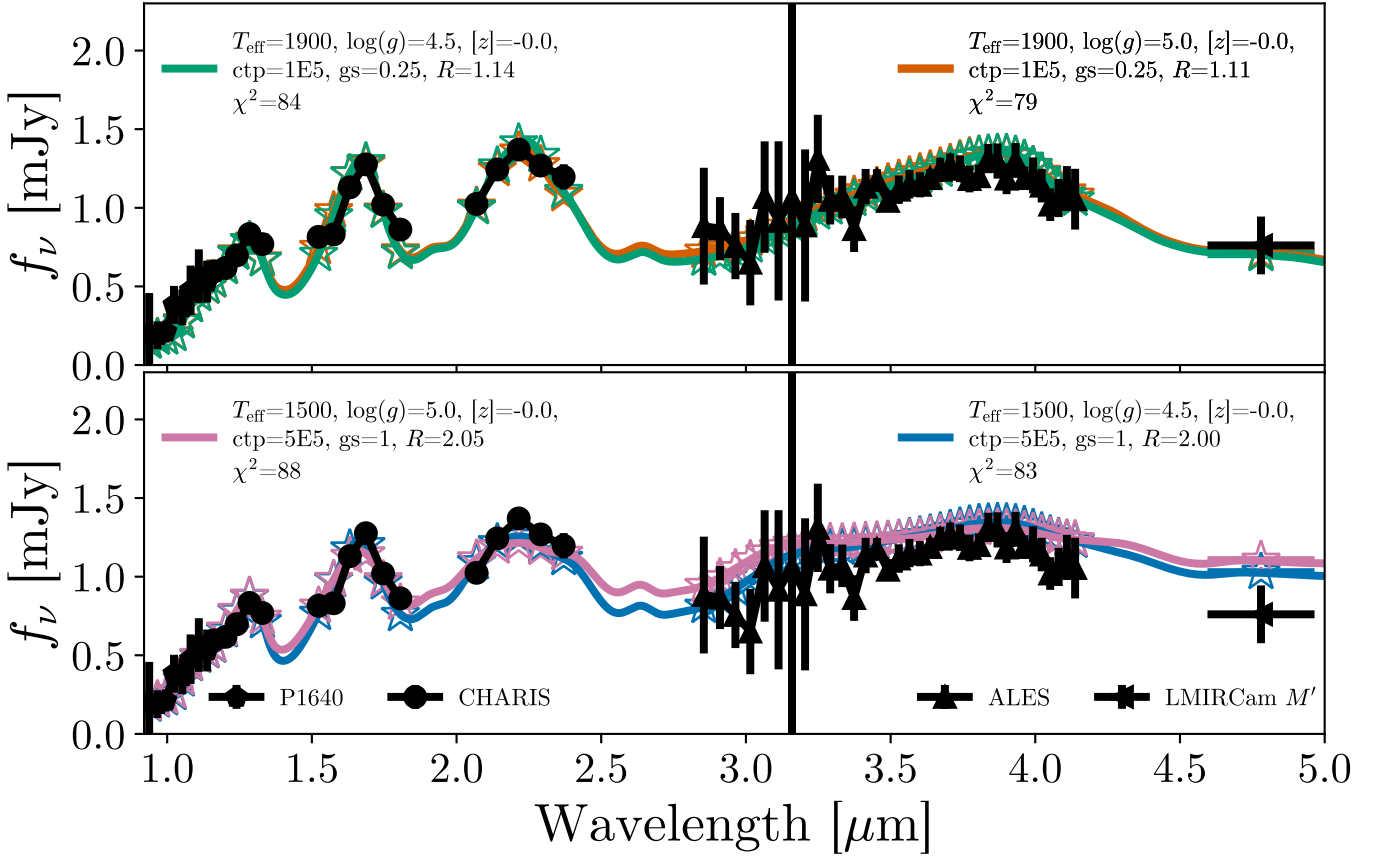}
\caption{Synthetic atmosphere model spectra compared to the 0.9 to 4.8 micron
SED of $\kappa$~And~b with the $L$-band flux scaling consistent with ALES and
LMIRCam constraints, $m_{L^{\prime}}=13.3$. All models listed in Table
\ref{modelRanges} with $>5\%$ likelihood are shown. Upper panel: Two models
with $T_{\mathrm{eff}}=1900$~K, cloud top pressures (ctp) of $10^{5}$ dyne
cm$^{-2}$, and 0.25 $\mu$m median grain size are allowed. Lower Panel: Two
models with $T_{\mathrm{eff}}=1500$~K and the lowest pressure cloud tops and
smallest grain size explored are allowed. Star symbols on each model atmosphere
indicate sampled fluxes used by the fitting routine.
\label{ALESSynthFits}} 
\end{figure*}

\subsection{High-Contrast Spectral Extraction} 
The superb performance of the LBT-AO system and the sensitivity of ALES within
LMIRCam combined to reveal $\kappa$~And~b in each of our 16 individual cubes.
As a result, aggressive software post-processing to separate the companion from
the primary star was not necessary, and we avoided all but the most simple
approach in order to minimize the introduction of hard-to-calibrate
spectrophotometric biases \citep[e.g.,][]{Lafreniere2007, Pueyo2016}. Working
on each wavelength slice independently (no spectral differential imaging), we
applied an unsharp mask high-pass spatial filter using a gaussian smoothing
kernal with $\sigma=2$ pixels (52 milliarcsecond) and then a basic angular
differential imaging (ADI) algorithm.  Our ADI approach involved
subtracting from each image at wavelength $\lambda$ the median of all other
images with wavelength $\lambda$. The closest any two images were in
parallactic angle was $3^{\degr}$, implying a minimum displacement of
$\sim\frac{\lambda}{\mathrm{D}}$ at the separation of the companion. The final
processed images for each wavelength are summed for presentation in Figure
\ref{cubestack}.

When using ADI-based image processing algorithms, extra care must be given to
photometric measurements to avoid biases related to source self-subtraction.
We injected a scaled negative version of the wavelength-dependent PSF into each
image slice at the position of the companion \citep[e.g.,][]{Lafreniere2007,
Pueyo2016}.  The scale factor and source position were fitted simultaneously
using the residuals after ADI processing to define the goodness of fit.  For
this purpose, we used the unsaturated data cube of $\kappa$~And~A for our
wavelength-specific PSFs.  With this approach, our extracted spectrum is
automatically corrected for telluric absorption, since it is present in both
the PSF and companion. To flux calibrate our extracted spectrum we multiplied
by a NEXTGEN \citep{Hauschildt1999} model A0 spectrum appropriately scaled to
yield the observed $L^{\prime}$ flux of $\kappa$~And~A
\citep[$m_{L^{\prime}}=4.32$][]{Bonnefoy2014}. An A0 star model is a close
approximation to the B9 spectral type of $\kappa$~And~A at these wavelengths.

\begin{figure*}
\includegraphics[width=\linewidth]{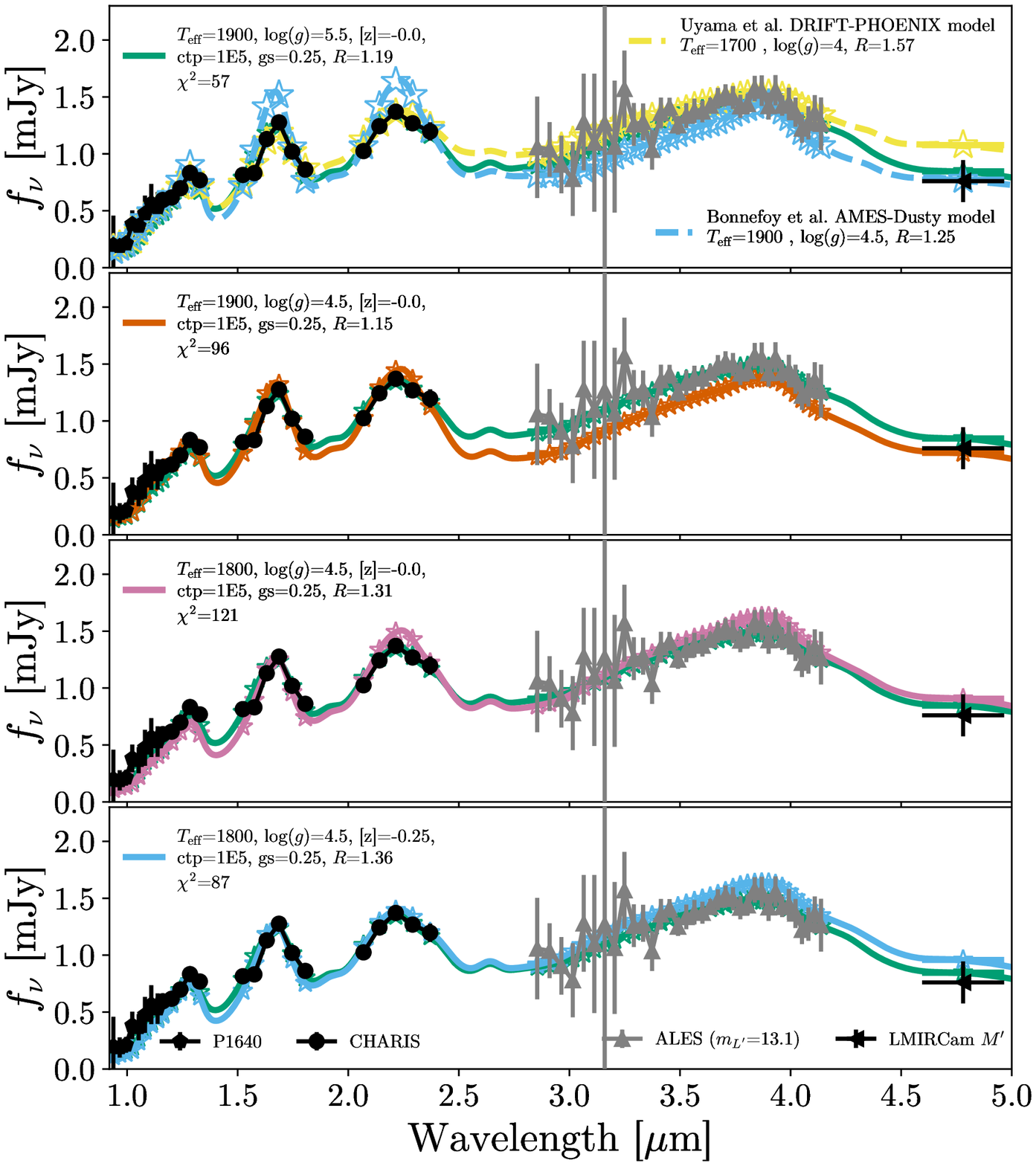}
\caption{Synthetic atmosphere model spectra compared to the 0.9 to 4.8 micron
SED of $\kappa$~And~b with the $L$-band flux scaling consistent with Subaru and
Keck photometry. Star symbols on synthetic spectra indicate sampled fluxes
compared to the observed data. Model atmosphere $T_{\mathrm{eff}}$ in Kelvin,
surface gravity, $g$, in cm s$^{-2}$, cloud top pressure, ctp, in dyne
cm$^{-2}$, median cloud particle size, gs, in microns, and object radius, $R$,
in $R_{\mathrm{Jup}}$. Our best fit model from the set indicated in Table
\ref{modelRanges} (solid teal green curve) appears in each panel. Upper Panel:
the best fit models reported by \citet{Bonnefoy2014} and \citet{Uyama2020} are
shown with dashed curves. These are not fits to the data shown, but are simply
taken from previous works and plotted. Second from top: The best-fit model has
high surface gravity that is unphysical. A model with surface gravity more
consistent with the predictions of evolutionary models is shown, but it is too
blue in its near-IR to L-band colors. Third from top: A cooler low-gravity
model better matches the gross colors of the data, but misses the shape of the
K-band peak. Bottom: A cooler low-gravity model with sub-solar metallicity
provides a plausible match to the data.  \label{KeckSynthFit}} \end{figure*}

Photometric uncertainty was estimated using the bootstrap method, repeating our
extraction and calibration procedure 30 times, each time selecting a different
random sample (with replacement) of our 16 frames \citep{Press2002}. We provide
our extracted spectrum in Table \ref{spectrum}.
\begin{deluxetable}{ccc}
\tablecolumns{3}
\tablewidth{0pt}
\tablecaption{ALES Spectrum of $\kappa$~And~b\label{spectrum}}
\tablehead{
    \colhead{Wavelength}& 
    \colhead{$F_{\nu}$}&
    \colhead{$\sigma_{F_\nu}$}\\
    \colhead{[$\mu$m]}&
    \colhead{[mJy]} &
    \colhead{[mJy]}
    }
\startdata
2.85 &0.88 &0.37\\
2.91 &0.87 &0.20\\
2.96 &0.76 &0.21\\
3.02 &0.65 &0.27\\
3.06 &1.07 &0.35\\
3.11 &0.92 &0.51\\
3.16 &1.06 &1.40\\
3.20 &0.89 &0.48\\
3.25 &1.31 &0.28\\
3.29 &1.05 &0.16\\
3.33 &1.07 &0.14\\
3.37 &0.86 &0.15\\
3.41 &1.13 &0.11\\
3.45 &1.18 &0.07\\
3.49 &1.04 &0.07\\
3.53 &1.12 &0.09\\
3.57 &1.15 &0.09\\
3.60 &1.15 &0.08\\
3.64 &1.19 &0.06\\
3.67 &1.24 &0.08\\
3.71 &1.27 &0.07\\
3.74 &1.25 &0.08\\
3.77 &1.18 &0.08\\
3.81 &1.20 &0.07\\
3.84 &1.31 &0.09\\
3.87 &1.30 &0.11\\
3.90 &1.19 &0.10\\
3.93 &1.30 &0.11\\
3.96 &1.19 &0.07\\
3.99 &1.21 &0.11\\
4.02 &1.14 &0.11\\
4.05 &1.02 &0.10\\
4.08 &1.06 &0.12\\
4.11 &1.14 &0.13\\
4.14 &1.05 &0.19
\enddata 
\end{deluxetable}

\section{Analysis} \label{analysisSec} 

\subsection{Tension with Earlier $L^{\prime}$ measurements} 

As a consistency check, we compared the magnitude of $\kappa$~And~b implied by
our flux-calibrated spectrum to previously reported photometry.  To do this, we
used the NIRC2 Lp filter curve to compare our measurement to that of
\citet{Bonnefoy2014}, revealing a $2\sigma$, 0.2 mag, discrepancy ---the ALES
spectrum (with $m_{L^{\prime}}=13.32\pm0.07$) being fainter than the reported
NIRC2 measurement (with $m_{L^{\prime}}=13.1$). The NIRC2 measurement reported by
\citet{Bonnefoy2014} is consistent with the IRCS measurement of $\kappa$~And~b
reported by \citet{Carson2013}. For an additional check, we compared to an
unpublished LBTI/LMIRCam $L^{\prime}$ observation of $\kappa$~And collected on
UT 2013 October 24 as part of the LEECH survey \citep{Stone2018}. The data
cover $97^{\degr}$ of parallactic rotation and were collected using only the
right-side 8.4 m mirror of the LBT.  Unsaturated images of the primary star
were obtained using a neutral density filter at the end of the observing block.
Preprocessing of the LMIRCam data followed the steps outlined in
\citet{Stone2018}, and high-contrast photometry was carried out exactly as
described above for a single wavelength slice of an ALES cube.  The LMIRCam
data yield $m_{L{\prime}}=13.36$, consistent with the ALES spectrum.  Thus, NIRC2
and IRCS suggest $\kappa$~And~b has $m_{L^{\prime}}=13.1$ whereas LMIRCam imaging and
ALES give $m_{L^{\prime}}=13.3$.

Since all four datasets were taken at different times, we address whether
variability could play a significant role in the 20\% flux discrepancy among
the measurements, it cannot.  Low-gravity early L-type objects typically vary
by $\lesssim1\%$ \citep{Metchev2015, Vos2020}.  Even the most extreme L-type
variables exhibit only $\sim5\%$ photospheric variability in the thermal-IR,
and these are of later spectral type than $\kappa$~And~b \citep[see][]{Zhou2020}.

The most conspicuous difference between the earlier $L^{\prime}$ photometry and
the measurements reported here, is that we used $\kappa$~And~A as photometric
calibrator for $\kappa$~And~b, while \citet{Carson2013} and
\citet{Bonnefoy2014} both use HR~8799~A.  We take $m_{L^{\prime}}=4.32$ for
$\kappa$~And~A from \citet{Bonnefoy2014}. Another significant difference is
that we used negative source injections to account for the throughput of our
high-contrast image processing approach, while both \citet{Carson2013} and
\citet{Bonnefoy2014} use aperture photometry corrected with a throughput
factor. Additional $L^{\prime}$ photometry of the system, including of the
primary, is warranted for a more accurate $L^{\prime}$ flux scaling of
$\kappa$~And~b. 

While the photometry differs by only $2\sigma$, the difference is large enough
to affect the best-fit atmospheric model parameters (see below). Therefore, we
elect to fit models to the SED of $\kappa$~And~b
using both the \citet{Carson2013} value ($m_{L^{\prime}}=13.1$) and the ALES
value ($m_{L^{\prime}}=13.3$).

\subsection{Fits to Brown Dwarf Template Spectra} \label{templateSec} We
compiled a set of brown dwarf spectra, covering spectral types M9 through T5,
to compare to $\kappa$~And~b. The brown dwarf data come from the Spex
instrument and are collected from \citet{Cushing2008} and \citet{Rayner2009}.
The selected spectra were chosen for their wavelength coverage, extending to
$4.1~\mu$m. We do not have access to a large number of objects with this broad
wavelength coverage. Future $L$- and $M$-band spectroscopy of brown dwarfs,
covering a wide range of spectral types and surface gravities, will be
important for facilitating empirical comparisons to the spectra of directly
imaged companions both for future ALES datasets and in the era of similar more
powerful instruments on giant segmented mirror telescopes \citep[e.g., METIS
and PSI-RED,][]{Brandl2014, Skemer2018b}. The James Webb Space Telescope will
play an important role in delivering empirical brown dwarf spectra spanning
near to mid infrared wavelengths.

For the comparison, we combined our new L-band ALES spectrum of $\kappa$~And~b
with the y-band spectrum from P1640 \citep{Hinkley2013}, and the J-, H-, and
K-band spectrum from CHARIS \citep{Currie2018}. We flux calibrate the y-band
P1640 spectrum using the y-band photometry reported in \citet{Uyama2020}. 

To fit the template spectra to the composite $0.9-4.1~\mu$m integral field
spectrograph (IFS) spectrum of $\kappa$~And~b, we first smoothed the Spex
spectra to match the resolving power of the three integral field spectrographs,
$R\sim20$. We propagated the noise in the Spex spectra through this process by
sampling the (assumed gaussian) noise of each data point and repeating the
smoothing process 100 times. The standard deviation of the resulting spread in
the R=20 Spex spectra was taken as the error.

We then used a $\chi^{2}$ fitting approach to find the best match, using
\begin{equation}\label{chiEquation} 
\chi^{2} = (f - sM)^{T} \Sigma^{-1} (f - sM), 
\end{equation} 
where $f$ is the observed spectrum, $M$ is the template
spectrum, $s$ is a scale factor to accommodate sources at different distances
and with different radii, and $\Sigma$ is the data covariance matrix. We
construct $\Sigma$ as a block diagonal matrix, with a block encoding the P1640
covariance, a block encoding the CHARIS covariance, and a block encoding the
ALES covariance. For ALES, we determine the covariance using the method of
\citet{Greco2016}, deriving $A_{\rho}=0.1$, $\sigma_{\rho}=0.16$,
$A_{\lambda}=0.25$, $\sigma_{\lambda}=0.01$, and $A_{\delta}=0.64$, with
variables as defined by those authors. We take the CHARIS block to be diagonal,
which is a good approximation for these data \citep{Currie2018}. The P1640 data
are not as aggressively binned as the CHARIS data, and like the ALES spectrum,
neighboring spectral channels appear correlated. \citet{Hinkley2013} do not
provide a measured covariance matrix for the P1640 data, so in order to down
weight the correlated data points from the P1640 spectrum, we use the same
empirical correlation parameters as determined for the ALES spectrum and apply
them to the P1640 data to create a covariance matrix. Using this P1640
covariance matrix did not change the resulting fit compared to assuming
uncorrelated error within the P1640 block.  Finally, we added to the diagonal
of $\Sigma$ the variance due to uncertainty in the template spectra.

We performed the fit for three scenarios: 1) Assuming $m_{L^{\prime}}=13.3$, as
measured by ALES and LMIRCam; 2) Assuming $m_{L^{\prime}}=13.1$, as measured by
the IRCS and NIRC2 instruments; and 3) for the $m_{L^{\prime}}=13.1$ case, in
addition to fitting for the best template, we also fit for the best reddening
parameters using the model of \citet{Cardelli1989} to modify each template.
This last method is motivated by \citet{Hiranaka2016} who show that high
altitude hazes in the atmospheres of brown dwarfs can mimic the effect of
interstellar reddening on the emergent spectrum.

In Figure \ref{SpexChi} we show how the resulting $\chi^{2}$ varies as
a function of spectral type for each scenario. In Figure \ref{SpexSpec} we show
the best fitting template spectrum for all cases. The best fit L3-type object,
2MASS~J15065441+1321060, is the same for each case and is also the best fitting
field-gravity template found by \citet{Uyama2020} who fit y-band photometry and
the CHARIS near-IR spectroscopy. \citet{Cushing2008} fit synthetic models to
the spectrum of 2MASS~J15065441+1321060 and report $T_{\mathrm{eff}}=1800$~K
and $\log(g)=4.5$.

The fit is best for the $m_{L^{\prime}}=13.3$ case, the SED of $\kappa$~And~b
appearing consistent with a typical field L3. The most significant discrepancy
between the high-contrast companion and the brown dwarf spectra is seen in the
last few spectral channels at the longest wavelengths. These wavelengths, near
the edge of the atmospheric transmission window can be affected by observing
conditions.

For the case of $m_{L^{\prime}}=13.1$, the near-IR to thermal-IR color is too
red.  Applying an interstellar reddening prescription \citep{Cardelli1989} can
improve the fit some, but the fitter is not at liberty to choose a strong
enough reddening to accommodate the thermal-IR to near-IR colors because such
a large value would change the ratios between the near-IR bands, which are
very-precisely constrained by the CHARIS data. Red near-IR to thermal-IR color
in L-dwarf spectra is known to correlate with low surface gravity
\citep[e.g.,][]{Filippazzo2015}.

\subsection{Fits to Model Atmospheres}
\begin{deluxetable*}{lccccc}
\tabletypesize{\footnotesize}
\tablecolumns{6}
\tablewidth{0pt}
\tablecaption{Atmosphere Model Fits to $\kappa$~And~b from the
literature\label{modelsSummary}}
\tablehead{
    \colhead{Model}& 
    \colhead{Spectral}&
    \colhead{$T_{\mathrm{eff}}$}&
    \colhead{$\log(g)$}&
    \colhead{Radius}&
    \colhead{Reference}\\
    \colhead{Description} &
    \colhead{Coverage} &
    \colhead{}&
    \colhead{}&
    \colhead{$R_{\mathrm{Jup}}$}&
    \colhead{}
    }
\startdata
DRIFT-PHOENIX    & 0.97-4.78 $\mu$m & 1700~K  & 4.0  & 1.57    & \citet{Uyama2020} \\
AMES-DUSTY       & 1.25-4.78 $\mu$m & 1900~K  & 4.5  & 1.25    & \citet{Bonnefoy2014} \\
\citet{Rice2010} & 0.9-1.32  $\mu$m & 2096~K  & 4.65 & \nodata & \citet{Hinkley2013} \\
\citet{Rice2010} & 1.47-1.78 $\mu$m & 1550~K  &      & \nodata & \citet{Hinkley2013} \\
\citet{Rice2010} & 0.9-1.78  $\mu$m & 2040~K  & 4.33 & \nodata & \citet{Hinkley2013} \\
\enddata 
\end{deluxetable*}

We fit the SED of $\kappa$~And~b to model spectra to better understand the
physical nature of the atmosphere of the high-contrast companion. Many previous
studies fit synthetic spectra to observations of $\kappa$~And~b using
a variety of observational constraints covering different wavelength ranges
and a variety of model atmosphere implementations. Table \ref{modelsSummary}
summarizes the best fit models from previous studies found in the literature.
Best-fit model atmosphere parameters span a temperature range consistent with
spectral type determined by comparison to brown dwarf spectra.

For the purposes of fitting synthetic models, we use the IFS spectra covering
0.9 to 4.1 $\mu$m described in Section \ref{templateSec} and combine with
  LBTI/LMIRCam $M^{\prime}$-band photometry from \citet{Bonnefoy2014}. We also
expand the covariance matrix $\Sigma$ by one dimension and add a new 1x1 block
to the diagonal to account for the measurement uncertainty in the
$M^{\prime}$-band flux. We then use Equation \ref{chiEquation} as the goodness of
fit metric to identify the best model parameters as constrained by the data. In
this case we take the scale factor
\begin{equation}
s = (Rq)^{2},
\end{equation}
where $R$ is the radius of the object (to be fit), and $q$ is the parallax of
the $\kappa$~And system, taken to be 19.98 milliarcseconds
\citep{GaiaCollaboration2018}.

The models we use are an extension of those described in \citet{Barman2011b},
\citet{Barman2015}, and \citet{Miles2018}.  Specifically, we can tune both the
cloud-top pressure (below which, climbing to higher altitudes, cloud particles
decay exponentially) and the median particle size within clouds.  The models
used here reach even lower cloud top pressures and smaller median particle
sizes than previous studies.  As we show below, these parameters appear
important for fitting the observations of young, cloudy L-dwarfs. Table
\ref{modelRanges} provides the parameter ranges of the models we used to fit
our data.

\begin{deluxetable*}{rllll}
\tabletypesize{\footnotesize}
\tablecolumns{5}
\tablewidth{0pt}
\tablecaption{Model Parameter Ranges\label{modelRanges}}
\tablehead{
    \colhead{$T_{\mathrm{eff}}$}& 
    \colhead{$\log(g)$}&
    \colhead{Cloud top pressure}&
    \colhead{Median particle size}&
    \colhead{Note}\\
    \colhead{K}& 
    \colhead{$g$ in ${\mathrm{cm~s}^{-2}}$}&
    \colhead{dyne cm$^{-2}$}&
    \colhead{$\mu$m}&
    \colhead{}
    }
\startdata
800 - 2100   & 3.5-5.5        & 5E5, 1E6, 4E6 & 1            & 100 K steps in $T_{\mathrm{eff}}$, 0.5 steps in $\log(g)$\\
1000 - 2000  & 4.75, 5.0, 5.5 & 1E7, 2E7, 3E7 & 0.25, 0.5, 1 & 100 K steps in $T_{\mathrm{eff}}$\\
1700 - 2000  & 4.0-5.5        & 1E5, 5E5      & 0.25         & 100 K steps in $T_{\mathrm{eff}}$, 0.5 steps in $\log(g)$\\
\enddata 
\end{deluxetable*}

In Figure \ref{ALESSynthFits} we show all the models allowed at the 95\%
confidence level for the case where the ALES spectrum corresponds to
$m_{L^{\prime}}=13.3$. Taking $\Delta\chi^2$ to be $\chi^2$-distributed with
five degrees of freedom (the four atmospheric model parameters in Table
\ref{modelRanges} and the object radius, which is simultaneously fit), these
are all the models with $\Delta\chi^2<11.3$.

We see the best-fit models break into two categories, a set with
$T_{\mathrm{eff}}=1900$~K, and a set with $T_{\mathrm{eff}}=1500$~K. This is
reminiscent of the multi-modal $\chi^{2}$-surfaces seen in previous fits to
early L-type objects \citep[e.g.,][]{Stone2016b}.

In Figure \ref{KeckSynthFit} we show the best fit model atmosphere for the case
where we scale the ALES spectrum to provide $m_{L^{\prime}}=13.1$. For this
case, only one model is allowed at 95\% confidence, having
$T_{\mathrm{eff}}=1900$~K, $\log(g)=5.5$, high-altitude clouds (cloud top
pressure of $10^{5}$ dyne cm$^{-2}$), and small grain sizes (0.25 $\mu$m
median).

In both Figures \ref{ALESSynthFits} and \ref{KeckSynthFit}, error bars on the
CHARIS data \citep[taken from][]{Currie2018} are smaller than the plotting
symbols. These tiny error bars strongly influence the fit; small changes in the
shape of the near-IR spectrum can drive huge changes in the $\chi^{2}$.  The
result is a narrow range of parameters formally allowed by the data. For
example, Figure \ref{ALESSynthFits} shows only four models within
$\Delta\chi^2$ of 11.3 from the best fit model ---the 95\% confidence range for
a model with 5 parameters. However, if we scale up the CHARIS errors to have
the same average fractional variance as the ALES spectrum, then the fitter
allows 19 models within 95\% confidence, spanning $T_{\mathrm{eff}}=1500$~K to
2000~K, with gravities ranging from $\log(g)=4.0$~to 5.5.

\begin{figure*}
\includegraphics[width=\linewidth]{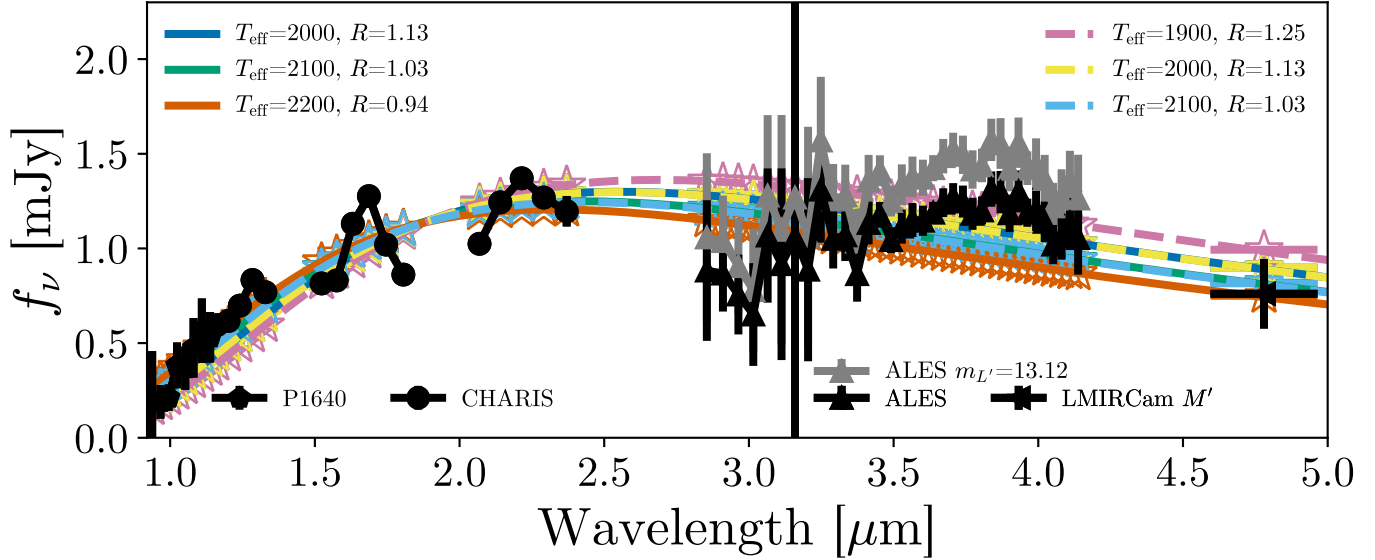} \caption{Blackbody
model fits to the SED of $\kappa$~And~b. ALES spectrum in black corresponds to
$m_{L^{\prime}}=13.3$, and solid blackbody curves are fit to the SED including
these data. ALES spectrum in gray corresponds to $m_{L^{\prime}}=13.1$, and
dashed blackbody curves are fit to the SED including these data.\label{BBFit}}
\end{figure*}

In all cases, clouds extending to very low pressures (high-altitudes) are
preferred by the fit. As cloud extent ---and opacity--- increases, molecular
bands in the emergent spectrum are muted and the emission trends toward
a blackbody shape. For example, \citet{Morzinski2015} showed for $\beta$~Pic~b
that fitting a blackbody spectrum to photometry covering the 0.99-4.8~$\mu$m
SED of the $T_{\mathrm{eff}}\sim1700$~K planet provides a better fit than
multiple more sophisticated atmosphere models. We fit blackbody models to the
measured portino of the $\kappa$~And~b SED, as shown in Figure \ref{BBFit}.
Models spanning temperatures from 1900~K to 2200~K provide good fits (95\%
confidence interval), with slightly cooler, but overlapping, temperatures for
the $m_{L^{\prime}}=13.1$ case. The blackbody implied radii range from 0.94 to 1.25
$R_{\mathrm{Jup}}$. For $\kappa$~And~b, the blackbody models always yield worse
fits than the atmosphere models described above, partly demonstrating the
strength of spectroscopy over photometry (the near-IR water bands are clearly
seen in the CHARIS data), and partly indicating the good performance of our
cloudy models.  

\subsubsection{Estimating $L_{\mathrm{bol}}$}\label{LbolSec}

Bolometric luminosity can be constrained using measurements of the SED covering
a broad wavelength range. With enough measurements, luminosity estimates are
robust to the choice of atmospheric model used to interpolate between and
extrapolate beyond the measured portions of the SED, yielding a robust value
(see Table \ref{LbolSec}). 

We combine the 0.9 to 4.8 $\mu$m measurements of $\kappa$~And~b together with
the well-fit (95\%-confidence) model atmospheres to estimate the bolometric
luminosity of the object. To do this, we used the synthetic atmosphere models
to extrapolate the SED to shorter and longer wavelengths and to interpolate
between the measured bands. We then integrated the semi-empirical SED and used
the $Gaia$ parallax to calculate the luminosity. To account for measurement
uncertainties we calculated the luminosity 20 times, each time sampling
$\Sigma$, the data covariance matrix, to modify the measured values, rescaling
the atmosphere model to fit each realization. The results are presented in
\ref{LbolTable}. We can estimate the scale of systematic error due to the
choice of atmosphere model and L-band flux scaling by noting the range of
luminosities measured for all cases. We determine that the luminosity of
$\kappa$~And~b is in the range
$\log_{10}(\frac{L_{\mathrm{bol}}}{L_{\odot}})=-3.69$~to~$-3.78$

\begin{deluxetable}{lcc}
\tabletypesize{\footnotesize}
\tablecolumns{3}
\tablewidth{0pt}
\tablecaption{$L_{\mathrm{bol}}$ Estimates for $\kappa$~And~b\label{LbolTable}}
\tablehead{
    \colhead{Model\tablenotemark{a}}& 
    \colhead{[$L_{bol}]$}&
    \colhead{\%-Measured\tablenotemark{b}}
    }
\startdata
\sidehead{$m_{L^{\prime}}=13.3$ Synthetic Atmosphere Models}
(1900, 5.0, 1E5, 0.25, $1.11\pm0.01$) & $-3.78\pm0.01$  & 63\%  \\
(1900, 4.5, 1E5, 0.25, $1.14\pm0.01$) & $-3.78\pm0.01$  & 64\%  \\
(1500, 5.0, 5E5, 1, $2.05\pm0.02$)    & $-3.71\pm0.01$  & 54\%  \\
(1500, 4.5, 5E5, 1, $2.00\pm0.02$)    & $-3.71\pm0.01$  & 57\%  \\
\sidehead{$m_{L^{\prime}}=13.3$ Blackbody Models}
(2000, $1.13\pm0.01$)    & $-3.71\pm0.01$ & 53\%  \\
(2100, $1.02\pm0.01$)    & $-3.71\pm0.01$ & 53\%  \\
(2200, $0.95\pm0.01$)    & $-3.0\pm0.01$ & 53\%  \\
\sidehead{$m_{L^{\prime}}=13.1$ Synthetic Atmosphere Models}
(1900, 5.5, 1E5, 0.25, $1.19\pm0.02$) & $-3.74\pm0.01$  & 61\%  \\
\sidehead{$m_{L^{\prime}}=13.1$ Blackbody Models}
(1900, $1.26\pm0.01$)    & $-3.69\pm0.01$ & 54\%   \\
(2000, $1.13\pm0.01$)    & $-3.69\pm0.01$ & 54\%  \\
(2100, $1.04\pm0.01$)    & $-3.69\pm0.01$ & 54\%  \\
\enddata 
\tablenotetext{a}{Synthetic atmosphere models are represented by\\
($T_{\mathrm{eff}}$, $\log(g)$, cloud top pressure, grain size, radius).}
\tablenotetext{b}{The fraction of the bolometric luminosity represented by the
portion of the SED with measured constraints.}
\end{deluxetable}

\section{Discussion} \label{discSec} 
Evolutionary models provide a way to evaluate the quality of atmospheric fits
and to distinguish physically reasonable atmospheric parameters from fits that
are hard to square with our understanding of the early evolution of substellar
objects.

The behavior of evolutionary models at young ages ($\lesssim100$~Myr) depends
sensitively on assumptions about the initial entropy of objects
\citep{Marley2007} but not on the choice of atmospheric model used to calculate
surface energy losses. Consequently, once initial entropy and age are fixed,
parameters predicted by evolutionary models, such as luminosity and radius (and
derivative quantities like effective temperature and surface gravity) are
robust ---modulo additional parameters effecting the internal physics of the
body, such as composition. 

For $\kappa$~And~b we fix initial entropy by considering hot-start models
exclusively.  Hot-start evolutionary models are consistent with the luminosity
and age-range of $\kappa$~And~b \citep{Bonnefoy2014} and are theoretically
supported given the large accretion rates required to build
a $\sim20~M_{\mathrm{Jup}}$ object during the lifetime of a typical B-star
protostellar disk \citep[e.g.,][]{Mordasini2013, Cumming2018}.

The results of \citet{Bell2015} and \citet{Jones2016} suggest an age of $\sim
50$~Myr for $\kappa$~And. We choose a conservative age range of 10-100 Myr, and
use the luminosity constraints shown in Table \ref{LbolTable} to compare to the
hot-start solar metallicity evolutionary model predictions of
\citet{Chabrier2000}. Our estimate of $L_{\mathrm{bol}}$ is
$\log_{10}(\frac{L}{L_{\odot}})=-3.69$ to $-3.78$, consistent with the value
reported by \citet{Bonnefoy2014}. 

We take a very conservative approach in comparing to the evolutionary model. We
do not interpolate the evolutionary model between mass bins. Rather, in Table
\ref{evolTable} we report the range of values for each predicted parameter for
the mass bins whose luminosities bracket the luminosity range of
$\kappa$~And~b.
\begin{deluxetable}{lcccc}
\tabletypesize{\footnotesize}
\tablecolumns{5}
\tablewidth{0pt}
\tablecaption{Evolutionary Model\tablenotemark{a} Predictions for
$\log_{10}(\frac{L_{\mathrm{bol}}}{L_{\odot}})=-3.69~\mathrm{to}~-3.78$\label{evolTable}}
\tablehead{
    \colhead{Age}& 
    \colhead{Mass}&
    \colhead{Radius}&
    \colhead{Temperature} &
    \colhead{$\log(\frac{g}{\mathrm{cm~s}^{-2}})$}\\
    \colhead{Myr}& 
    \colhead{$M_{\mathrm{Jup}}$}&
    \colhead{$R_{\mathrm{Jup}}$}&
    \colhead{K} &
    \colhead{}
    }
\startdata
10  & 9-10  & 1.51-1.53 & 1635-1731 & 4.01-4.04\\
40  & 12-15 & 1.36-1.61 & 1564-2025 & 4.17-4.23\\
50  & 20-25 & 1.37-1.39 & 1760-2004 & 4.44-4.53\\
70  & 20-30 & 1.30-1.32 & 1604-2050 & 4.49-4.65\\
100 & 25-40 & 1.23-1.29 & 1655-2234 & 4.63-4.80\\
\enddata 
\tablenotetext{a}{\citet{Chabrier2000}}
\end{deluxetable}

We conclude that the $T_{\mathrm{eff}}=1500$~K models in Figure
\ref{ALESSynthFits} are well outside expectations for an object with the
luminosity of $\kappa$~And~b and are inconsistent with the predictions of
evolutionary models. Table \ref{evolTable} shows a $T_{\mathrm{eff}}=1564$~K
prediction for an age of 40~Myr, but this is due to the coarseness of the
evolutionary model grid.  The luminosity corresponding to
$T_{\mathrm{eff}}=1564$~K is $\log_{10}(\frac{L}{L_{\odot}})=-3.99$, well below
what we measure.

For the atmosphere fits to the $m_{L^{\prime}}=13.3$ case, two 1900~K models
are allowed within the 95\% confidence interval, one with $\log(g)=4.5$ and one
with $\log(g)=5.0$. The $\log(g)=4.5$ model is consistent with the evolutionary
model ranges listed in Table \ref{evolTable}. Agreement between well-fit model
atmosphere parameters and evolutionary model predictions is not a given, and in
this case such agreement depends on specific cloud properties, including a very
low cloud top pressure ($10^{5}$ dyne cm$^{-2}$) and small median grain size
(0.25~$\mu$m). While our atmosphere model fits match the evolutionary model
predictions for both $T_{\mathrm{eff}}$ and $\log(g)$, the predicted radius
is larger than that inferred with an atmospheric model fit. This suggests that
a lower $T_{\mathrm{eff}}$ model, which would require a larger scaling (radius)
to intersect the data may be more appropriate. Both temperature and median
grainsize both effect the model atmospheres in a similar way, imposing a global
tilt across the 0.9 to 5 micron wavelength range. Consequently, finding
a larger radius lower-temperature model may require further tuning the cloud
parameters.

For the case with $m_{L^{\prime}}=13.1$ we find only one model atmosphere
within the 95\% confidence interval.  In the top panel of Figure
\ref{KeckSynthFit}, we show this model and compare to the best-fit models from
\citet{Bonnefoy2014} and \cite{Uyama2020}. The 1700~K DRIFT-PHOENIX model
identified by \citet{Uyama2020} provides a good overall fit and is consistent
with the predictions of the evolutionary model. Our model provides a better
overall fit to the data, especially providing a better match to the L-M color
of $\kappa$~And~b. However, our best fit model indicates a surface gravity of
$\log(g)=5.5$, inconsistent with the predictions of the evolutionary model.

We show a lower gravity fit to the data in the second panel from the top of
Figure \ref{KeckSynthFit}. This model is statistically ruled out by the fit
because it underpredicts the L-band flux. This is the model that is allowed in
the case of a fainter L-band flux discussed above. We note that lower-gravity
produces a bluer near-IR to thermal-IR color. This may seem
counterintuitive given that low-gravity brown dwarfs are known to be red for
their spectral type \citep[see][]{Faherty2016}. However, effective temperature
does not map to spectral type as readily for substellar objects as it does for
stars. Both \citet{Stephens2009} and \citet{Filippazzo2015} show red
low-gravity brown dwarfs are cooler than their higher gravity counterparts with
the same spectral type designation, and the discrepancy can be a few hundred
Kelvin. So ``low-gravity objects are red" is a statement that objects appear as
earlier type for a given $T_{\mathrm{eff}}$, rather than as extra red at fixed
$T_{\mathrm{eff}}$.

Since the low-gravity model is too blue a cooler atmosphere may provide
a better fit. We show in the third panel from the top of Figure
\ref{KeckSynthFit} an 1800~K model that appears to get the gross colors of the
SED correct. This model is statistically ruled out because of the
residuals at the peak of the K-band.

The peak of the K-band is particularly sensitive to the pressure of the
photosphere due to the significant opacity within the band from collisionally
induced absorption of molecular hydrogen. High-surface gravity is one way to
affect a high-pressure photosphere, another is to remove metals from the
atmosphere, making it more transparent and faciltating a view to deeper,
higher-pressure levels. In the bottom panel of Figure \ref{KeckSynthFit} we
show an 1800~K, $\log(g)=4.5$ model with subsolar metallicity. This model
provides a very good fit to the data and provides a radius consistent with
evolutionary models. Formally, with a $\Delta\chi^2=30$ from the best
fit model, the subsolar metallicity model is ruled out at the $4\sigma$ level
(six degrees of freedom). Remaining issues could likely be resolved with minor
modifications to the atmospheric parameters, but a detailed focused fit is
beyond the scope of this paper and is unwarranted given the level of systematic
errors present in the dataset.

Given the young age of the $\kappa$~And system it is unlikely to be
significantly metal poor. We looked up the members of the Columba association
listed in \citet{Bell2015}  in the Hypatia Catalog Database of stellar
abundances \citep{Hinkel2014}. We found four matches covering spectral types
spanning F3 to G3.  These four have median [Fe/H] spanning 0.16 to 0.97.  If
the atmosphere of $\kappa$~And~b is metal difficient compared to stellar
abundance, this is likely a signature of the formation process, and could be an
important clue to better understand the physical mechanism that produced the
low-mass companion. 

In both the $m_{L^{\prime}}=13.3$ and the $m_{L^{\prime}}=13.1$ cases, good
alignment with the data requires specific cloud properties, namely a low cloud
top pressure and small median grain sizes. This is a robust result, not
affected by choosing between the two $L^{\prime}$ photometric scalings. We note
that the DRIFT-PHOENIX model shown in Figure \ref{KeckSynthFit} also includes
a small average grain size in the upper layers of the atmosphere
\citep{Witte2009, Witte2011}.

\section{Conclusion} We present the first high-contrast $L$-band spectrum from
LBTI/ALES. The spectrum of $\kappa$~And~b ---a young low-mass companion with
mass ratio $\lesssim0.7\%$ with respect to its B9 host star--- is used to
constrain atmosphere models. Our ALES observation is about 20\% fainter over
the $L^{\prime}$ band compared to previoiusly published photometric
measurements.  A 2013 LBTI/LMIRCam observation of $\kappa$~And at $L^{\prime}$
yields a flux consistent with the ALES value. We combine ALES data with spectra
covering the Y to K bands and with $M^{\prime}$ photometry, yielding
measurements of the substellar companion SED spanning 0.9 to 4.8~$\mu$m. Using
the ALES $L$-band flux scaling, the data are well fit by an L3-type brown dwarf
from the field.  If previous photometry is more accurate, then none of the
template brown dwarf spectra we compare to can match the red near-IR to
thermal-IR color.  The data precisely constrain the bolometric luminosity of
the object, which we use as input to evolutionary models. We find atmospheric
models consistent with the predictions of evolutionary models for atmospheres
with $T_{\mathrm{eff}}=1800$ to $1900$~K and $\log(g)=4.5$ to $5$. Clouds
composed of small grains extending to high altitudes are required by the data
whether the ALES $L$-band flux scaling is used or not.  There is a hint of
substellar metallicity in the case of the brighter $L$-band flux scaling.
Future observations improving the $L^{\prime}$-band photometry of both
$\kappa$~And~A and b will help clarify if subsolar metallicity is required.
Improved atmospheric constraints will also be facilitated by independent
verification of the shape of the K-band emission and improved precision of the
$M^{\prime}$ photometry.

\acknowledgements This paper is based on work funded by NSF Grants 1608834,
1614320 and 1614492. Work conducted by Laci Brock and Travis Barman was also
        supported by the National Science Foundation under Award No. 1405504.
J.M.S. is supported by NASA through Hubble Fellowship grant HST-HF2-51398.001-A
awarded by the Space Telescope Science Institute, which is operated by the
Association of Universities for Research in Astronomy, Inc., for NASA, under
contract NAS5-26555.  Z.W.B. is supported by the National Science Foundation
Graduate Research Fellowship under Grant No. 1842400.  C.E.W also acknowledges
partial support from NASA grant 80NSSC19K0868. The LBT is an international
collaboration among institutions in the United States, Italy and Germany.  LBT
Corporation partners ar: The University of Arizona on behalf of the Arizona
university system; Istituto Nazionale di Astrofisica, Italy; LBT
Beteiligungsgesellschaft, Germany, representing the Max-Planck Society, the
Astrophysical Institute Potsdam, and Heidelberg University; The Ohio State
University, and The Research Corporation, on behalf of The University of Notre
Dame, University of Minnesota, and University of Virginia. We thank all LBTI
team members for their efforts that enabled this work.  This work benefited
from the Exoplanet Summer Program in the Other Worlds Laboratory (OWL) at the
University of California, Santa Cruz, a program funded by the Heising-Simons
Foundation.  This work has made use of data from the European Space Agency
(ESA) mission {\it Gaia} (\url{https://www.cosmos.esa.int/gaia}), processed by
the {\it Gaia} Data Processing and Analysis Consortium (DPAC,
\url{https://www.cosmos.esa.int/web/gaia/dpac/consortium}). Funding for the
DPAC has been provided by national institutions, in particular the institutions
participating in the {\it Gaia} Multilateral Agreement. The research shown here
acknowledges use of the Hypatia Catalog Database, an online compilation of
stellar abundance data as described in Hinkel et al. (2014, AJ, 148, 54), which
was supported by NASA's Nexus for Exoplanet System Science (NExSS) research
coordination network and the Vanderbilt Initiative in Data-Intensive
Astrophysics (VIDA).

\facilities{LBT (LBTI/LMIRCam, LBTI/ALES)}

\software{Astropy \citep{astropy2013},
Matplotlib \citep{matplotlib},
Scipy \citep{2020SciPy-NMeth},
SPLAT (https://github.com/aburgasser/splat),
MEAD \citep{Briesemeister2018}
}

\bibliographystyle{aasjournal}
\end{document}